# A Multiphysics Analysis and Investigation of Soft Magnetics Effect on IPMSM: Case Study Dynamometer

Ali Amini
*Department of Electrical Engineering*
*Iran University of Science and Technology*
Tehran, Iran
al_amini76@elec.iust.ac.ir

MohammadSadegh KhajueeZadeh
*Department of Electrical Engineering*
*Sharif University of Technology*
Tehran, Iran
mohammadkhajuee@yahoo.com

Abolfazl Vahedi
*Department of Electrical Engineering*
*Iran University of Science and Technology*
Tehran, Iran
avahedi@iust.ac.ir

*Abstract*— Nowadays, Interior Permanent Magnet Synchronous Motors (IPMSMs) are taken into attention in the in-dustry owing to their advantages. Moreover, in many cases, performing static tests is not enough, and in-vestigating electric machines under dynamic condi-tions is necessary. Accordingly, by employing a dy-namometer system, the dynamic behavior of the elec-tric machine under test is investigated. Among the dynamometers, the best is the Alternating (AC) dy-namometer because the basic dynamometers cannot take loads with high complexity. So, in the following study, two IPMSM with V-type and Delta-type rotor configurations are designed and suggested to employ in AC dynamometer. Any non-ideality in the electric machines of AC dynamometers, electrically and me-chanically, causes errors in the measurement of the motor under test. Electrically and mechanically, the behavior of a system significantly depends on the used soft magnetics besides its physical and magnetic configuration. Accordingly, by performing a Mul-tiphysics analysis and using the FEM tool to change the soft magnetics in the rotor and stator core, com-paring the electric motors' behavior in the AC dyna-mometer is investigated under the same operating conditions electrically and mechanically. Finally, which soft magnetics is more satisfactory for the AC dynamometer can be seen.

## I. INTRODUCTION

A dynamometer system is widely employed for testing electric and combustion machines. The exact and trustworthy investigations are necessary for improving the operation of the electric machine. In many cases, performing static tests is not enough, and studying the machines under dynamic tests is necessary. Accordingly, the dynamic operating of the machines under test will be investigated by employing of dynamometer system. Basic dynamometers such as Powder Brake (PB), eddy current brake, Hysteresis Brake (HB), and Water Brake (WB) can be employed. The significant challenge of the basic brakes is the inability to take dynamic loads, which is more obvious with the growth of electric motor usage and various loads in the industry. Therefore, owing to the above disadvantages, Alternating (AC) dynamometers are employed for the test bench. The AC dynamometer system consists of two electric machines, which have mechanical coupling to each other through the rotor shaft axis. According to Figure 1, one of these electric machines is for producing the load, which can be regarded as an AC dynamometer [1]. Recently, dynamometer control has been investigated [2]. On the contrary, no significant study on the employed electric machines in AC dynamometers. Due to availability and lower cost, Induction Motors (IMs) are more employed in AC dynamometers in the industry. On the other hand, using more efficient electric motors is taken into attention nowadays. According to the dynamometer operating, the designed motor must have low Torque Ripple ($T_{ripple}$) and subsequently low vibrations, high Maximum Torque ($T_{max}$), fast dynamic (low inertia), and high Average Torque ($T_{avg}$). Thus, the Interior Permanent Magnet Synchronous Motor (IPMSM) is one of the best motors for AC dynamometers [3]-[5]. Among the various type of rotor configurations in the IPMSMs, the V-type series are employed owing to the satisfactory $T_{ripple}$, $T_{max}$, and $T_{avg}$ [6]. Accordingly, two IPMSMs with V-type rotor configurations for the AC dynamometer are designed in the following study. Although Multiphysics analysis is attractive in electric motors, the effect of soft magnetics in the design and manufacturing of electric machines in the industry is not investigated appropriately and comprehensively based on usage. Also, the behavior of a system, mechanically and electrically, has a significant dependency on the employed soft magnetics besides its physical and magnetic configuration. Therefore, changing the soft magnetics in the rotor and stator core will yield various mechanical and electrical behaviors under the same operating conditions. Accordingly, it should be investigated which soft magnetics will be better to use in electric motors; high-cost materials with promising electrical advantages won't necessarily give satisfactory mechanical behaviors.

Consequently, in the following study, besides the design of two satisfactory electric machines for AC dynamometers, comparing and investigating the effect of soft magnetics on their behavior electrically and mechanically is regarded. Hence various soft magnetics were employed in the rotor and stator core of the test bench's electric machines for this investigation. Accordingly, in section II, comparing employed soft magnetics is investigated. Then in section III, two motors specifically for the AC dynamometer are designed, and the effect of soft magnet on the significant behavior of the AC dynamometer is investigated electrically and mechanically through Finite Element Analysis (FEA).

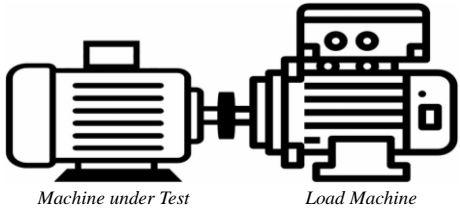

Figure 1. Schematic of test bench with employing AC dynamometer.

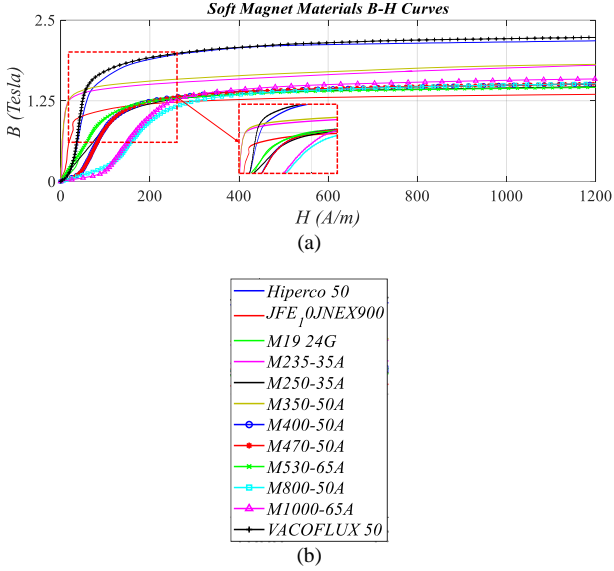

Figure 2. (a) B-H curves of the investigated soft magnetics. (b) Legend guide of curves in Figure2 and Figure 4.

## II. THE USED SOFT MAGNETIC MATERIALS

The use of soft magnetics in manufacturing electric machines is extremely regarded. The advantages of soft magnetics employed in electric machines such as high permeability for reducing the Magnetic Resistance ($R_{mag}$) of the circuit, high saturation flux density for minimizing the volume and weight of rotor and stator core, and low losses owing to the effect of losses on efficiency and increasing warmth [7]. Accordingly, comparing the used soft magnetics in IPMSM for AC dynamometer is investigated in this section. As displayed in Figure 2, the B-H curve can be used to investigate the high saturation flux density and high permeability. The B-H curve is generally used to show a nonlinearity in the magnetic behavior of ferromagnetic material in a magnetic field. The measuring B-H curve in the laboratory is by following the guidelines. However, the measurements cannot be directly in the above saturation called over fluxed regions. Accordingly, B-H curve data extraction in the over fluxed region is usually through using the Standard Extrapolation Methods (SEMs) such as Simultaneous Exponential Extrapolation (SEE) [8].

Additionally, the loss curve and the efficiency of motors can be employed to investigate the losses. This study investigates soft magnetics, including "Hiperco 50", "JFE_10JNEX900", "VACOFLUX 50", and "electrical steel M series alloys." According to Figure 2, the highest saturation flux density belongs to "VACOFLUX 50" and "Hiperco 50". The first alloy is one of the iron-cobalt alloys (CoFe) called "Permendor," which contains 2% vanadium, 49% cobalt, and 49% iron. Besides the high saturation flux density that is useful for reducing the volume and weight of the electric machines, this alloy is highly satisfactory for the high value of $T_{avg}$.

Additionally, the mechanical strength can be changed according to lower iron loss and higher magnetic permeability by adjusting the ratio of cobalt and iron. Also, Hiperco 50 alloy analogous to VACOFLUX 50 contains iron-cobalt (CoFe) alloys and also has high permeability, which makes it satisfactory for the high value of $T_{avg}$. CoFe alloys are costly, owing to a significant amount of cobalt. JFE-Steel materials are silicon alloys, which are highly satisfactory for high-frequency values. Consequently, besides being employed in manufacturing electric machines, they are widely employed in switching power supply devices and are increasingly employed in the power supply of Hybrid Electric Vehicles (HEVs) [9]. Generally, the most widely used soft magnet in electric machines is iron alloyed with silicon. Adding silicon will be useful for reducing the de-hysteresis force and increasing the stability of magnetic effects for a longer time. However, it causes disadvantages such as a slightly decreasing saturation flux density.

## III. COMPARISON AND INVESTIGATION

The electric machine employed in the AC dynamometer must have low inertia, fast dynamic, high efficiency, low $\boldsymbol{T_{ripple}}$, minimum Cogging Torque ($\boldsymbol{T_{cogging}}$), high value of $\boldsymbol{T_{max}}$ and $\boldsymbol{T_{avg}}$, and subsequently satisfactory vibration in faulty and stress conditions. Among the above, low $\boldsymbol{T_{ripple}}$ and minimum $\boldsymbol{T_{cogging}}$ are the most noteworthy, which a high value of $\boldsymbol{T_{ripple}}$ in dynamometer makes a significant vibration and error in motor under test. According to the above objectives, two IPMSMs with various magnet array types in rotor configuration are suggested and designed to employ in the AC dynamometer, which skewing the stator's slots by 5%, optimally determining the size of the RIBs, size of the magnets, the angle of the magnets, and the distance between the magnets' center and the rotor's center are regarded for reducing the $\boldsymbol{T_{ripple}}$ and increasing the $\boldsymbol{T_{avg}}$. Technical design data of the IPMSMs are given in Table 1, such as air gap of 0.48 mm and magnet width of 3 mm are analogous, excluding the double layer magnet in Delta-type rotor configuration. Additionally, both rotor and stator configurations are analogous in laminations' material. The topology of designed IPMSMs is shown in Figure 3. Based on the multiphysics analysis, the design of an electric machine must regard the effect of the employed soft magnetics optimally. Therefore, electrically and mechanically, the soft magnetics' effect on motor behavior has been investigated under the Maximum Torque Per Ampere (MTPA) control strategy.

### A. Electrical Analysis

High $T_{ripple}$ and low $T_{max}$ of electric motors in AC dynamometer will cause the behavior of the motor under test not show satisfactory. So, comparing the $T_{ripple}$, $T_{max}$, $T_{avg}$, and efficiency is shown in the electrical analysis by employing various soft magnetics.

Table 1
Technical design data of the investigated IPMSMs

| Designed Parameters | Unit | Value |
|---|---|---|
| Number of Phases | - | 3 |
| Number of Poles | - | 8 |
| Rated Speed | RPM | 3000 |
| Line Peak Current | A | 300 |
| Winding Layers | - | 1 |
| Parallel Paths | - | 2 |
| Stator Outer Diameter | mm | 193.42 |
| Stator Inner Diameter | mm | 132.4 |
| Stator Slots | - | 48 |
| Slot Depth | mm | 23.2 |
| Tooth Width | mm | 3.32 |
| Stator Stack Length | mm | 160 |
| Magnet | - | N30UH* |

*N30UH: It belongs to the group of NdFeB magnets.

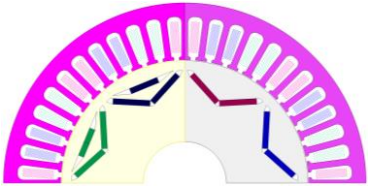

Figure 3. Topology of the designed and investigated IPMSMs: V-type (Right), Delta-type (Left).

*A.1. Torque Characteristics*
According to Table 2 and Figure 4, the lowest $T_{ripple}$ in the V-type IPMSM belongs to employing VACOFLUX 50, and in the Delta-type IPMSM belongs to employing M350-50A. The negative effect of VACOFLUX 50 and Hiperco 50 in the Delta-type rotor configuration on the $T_{ripple}$ is noteworthy, which shows that soft magnetics containing cobalt will intensify the $T_{ripple}$ of the Delta-type IPMSM rotor configuration. Increasing the magnetic transfer flux between the rotor and the stator can cause intensify the $T_{ripple}$ in Delta-type IPMSM. As shown in Figure 4(a), the highest value of $T_{avg}$ and $T_{max}$ in both topologies belong to Hiperco 50 material, although VACOFLUX 50 behaves satisfactorily too. As displayed in Figure 4(b), the lowest $T_{cogging}$ of both topologies belong to Hiperco 50 and M800-50A. Generally, the higher value of $T_{avg}$ and $T_{max}$ by soft magnetics containing cobalt is owing to their higher saturation flux density than the other soft magnetics [10].

*A.2. Efficiency*
High-efficiency electric machines play a key role in reducing electrical energy usage and improving system efficiency. Accordingly, the electric machine employed in the AC dynamometer must have high efficiency. Thus, the soft magnetics' effect on the efficiency of the designed topologies, including V-type and Delta-type IPMSM, is investigated. So, comparing the ratio of the losses in investigated soft magnetics is shown. The given efficiency values in Table 2 are written as below:

$$\eta = \frac{P_{out}}{P_{out} + P_c + P_{cu} + P_m} \quad (1)$$

Where, $P_c$ is the losses of rotor and stator cores, $P_{cu}$ is windings' losses, $P_m$ is PMs' losses, and $P_{out}$ is the output power of the designed motors. Hysteresis and eddy current losses have been regarded in $P_c$.

Table 2 shows that the highest efficiency of both topologies belongs to Hiperco 50 material, although M235-35A and M250-35A behave satisfactorily too. Nowadays, Hiperco 50 is at the center of attention, contrary to its high cost, owing to its high efficiency. The efficiency of both topologies with Hiperco 50 material is higher than 95%, which is a significant region. Consequently, it can be seen that Hiperco 50 material has a lower amount of loss in the loss curve to the flux density at the investigated frequency values. In the following, Hiperco 50, M800-50A, M235-35A, and VACOFLUX 50, which behaves more satisfactory than the others, will be analyzed.

Table 2
Outcomes of the designed and investigated IPMSMs with various soft magnetics in electrical analysis

| Soft Magnet | V type | | | Delta type | | |
|---|---|---|---|---|---|---|
| | Average Torque (Nm) | Torque Ripple (%) | System Efficiency (%) | Average Torque (Nm) | Torque Ripple (%) | System Efficiency (%) |
| Hiperco 50 | 121.59 | 8.339 | 95.27 | 118.94 | 10.422 | 95.201 |
| JFE_10JNEX900 | 98.81 | 8.7636 | 93.871 | 97.538 | 7.6399 | 93.883 |
| M19 24G | 103.2 | 8.0931 | 93.229 | 102.32 | 7.6974 | 93.313 |
| M235-35A | 103.41 | 8.6299 | 94.316 | 103.29 | 8.0746 | 94.349 |
| M250-35A | 103.71 | 8.7484 | 94.309 | 103.03 | 8.3417 | 94.321 |
| M350-50A | 101.09 | 7.8293 | 93.085 | 100.38 | 7.1591 | 93.165 |
| M400-50A | 101.34 | 9.072 | 92.792 | 99.479 | 7.7737 | 92.828 |
| M470-50A | 104.51 | 8.6362 | 93.136 | 103.88 | 8.2997 | 93.234 |
| M530-65A | 102.65 | 9.0422 | 92.982 | 101.14 | 8.1861 | 93.039 |
| M800-50A | 101.24 | 9.0122 | 92.296 | 99.67 | 7.6873 | 92.345 |
| M1000-65A | 102.85 | 8.6653 | 92.076 | 101.66 | 7.3952 | 92.18 |
| VACOFLUX 50 | 120.12 | 7.6989 | 93.862 | 118.19 | 9.7173 | 93.929 |

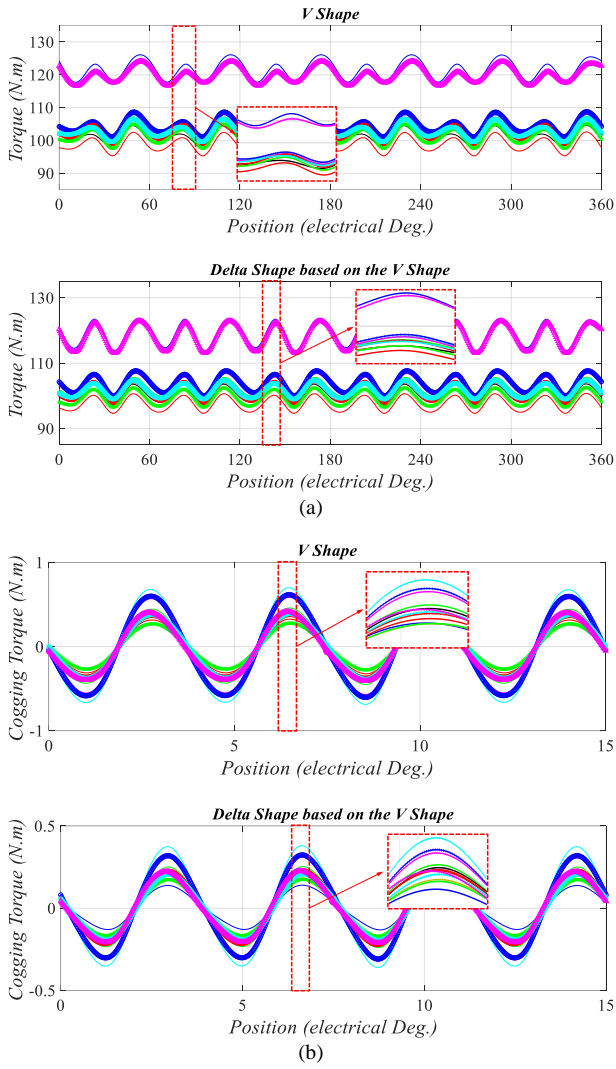

Figure 4. Torque diagrams of the investigated IPMSMs with various soft magnetics: (a) $T_{out}$, (b) $T_{cogging}$.

### B. Mechanical Analysis

Mechanical analysis is necessary for operating optimally and out of the fault in the electric machines of the AC dynamometer. Thus, the soft magnetics with better behavior in the electrical analysis are subjected to mechanical analysis. As the mechanical behavior of a system has a significant dependency on the employed materials besides the physical configuration, various mechanical behavior under the same operating conditions by changing the soft magnets in the rotor and stator cores will be yielded.

### B.1. Stress

Considering that the Output Torque ($T_{out}$) of the motor has non-ideality and oscillations, the forces will be yielded through solving the Dynamic Equations (DEqs) governing the systems, which can be in two modes, including shear stress or axial stress. Since the $T_{out}$ is obtained directly from the rotor and is the rotating part of the electric motor, all analyzes will be in the rotor. The stress in the non-linear region (Plastic region) after the yield stress is generally written as the ratio of the force to the applying area, as displayed in (2):

$$\tau = F/A \qquad (2)$$

Where $F$ is force, $A$ is applying area, and $\tau$ is stress. As shown in (3), the force has a dependency on the eccentricity percentage, besides the rotor angular velocity and the mass of the rotor, directly. So, with increasing angular velocity, the average stress value also grows exponentially [11].

$$F = me\omega^2 \qquad (3)$$

Where, $m$ is mass, $e$ is percentage of eccentricity, and $\omega$ is rotor angular velocity. Since always a bit percentage of eccentricity exists in rotor owing to the uncertainty in the manufacturing of the electric machines. So, the electric machine's force, stress, and vibrations in a healthy condition can be analyzed. Table 3 shows that although the rotor configuration is the same, due to the discrepancy in the motor's electrical operating, such as $T_{ripple}$ and the rotor's mass with various soft magnetics, the amount of force and stress on the rotor under the same operating conditions are various. Besides the non-linear region of stress, in the linear region, stress can be regarded as the ratio of the force on the applying area of it, according to (2), or the yield stress will be written as below in (4), with a dependency on Young's modulus of soft magnetics and tension value. So, various Young's modulus will give various yield stress, according to Table 4.

$$\sigma = E\varepsilon \qquad (4)$$

Where, $E$ is Young's modulus, $\varepsilon$ is tension, and $\sigma$ is yield stress [11]. According to Table 3, the lowest and highest average stress in both topologies belongs to M235-35A and VACOFLUX 50, respectively, in the same configuration and operating conditions, which is owing to the discrepancy in rotor mass and $T_{ripple}$ in various soft magnetics. Also, the average stress of the Delta-type is lower than the V-type rotor configuration.

Table 3
Outcomes of the designed and investigated IPMSMs with various soft magnetics in mechanical analysis

| Soft Magnet | V type | | | Delta type | | |
|---|---|---|---|---|---|---|
| | RLS* Avg (MPa) | Rotor Weight (Kg) | RLD Avg (mm) | RLS Avg (MPa) | Rotor Weight (Kg) | RLD Avg (mm) |
| **Hiperco 50** | 1.46 | 13.96 | 0.00033 | 1.07 | 13.67 | 0.00035 |
| **M235-35A** | 1.37 | 13.23 | 0.00035 | 1.01 | 12.98 | 0.00037 |
| **M800-50A** | 1.39 | 13.23 | 0.00031 | 1.02 | 12.98 | 0.00033 |
| **VACOFLUX 50** | 1.49 | 13.98 | 0.00028 | 1.08 | 13.68 | 0.00029 |

*Notation: RLS is Rotor Lamination Stress, RLD is Rotor Laminations Displacement.*

Table 4
Mechanical characteristics of the investigated soft magnetics

| Soft Magnet | Young's Coefficient (MPa) | Yield Stress (MPa) |
|---|---|---|
| Hiperco 50 (0.15mm) | 207000 | 393 |
| M235-35A | 185000 | 460 |
| M800-50A | 210000 | 300 |
| VACOFLUX 50 | 250000 | 390 |

Table 5
Lateral natural frequency values of the investigated IPMSMs' stator with various soft magnetics

| Mode | Lateral Natural Frequency | | | |
| | Hiperco 50 (0.15mm) | M235-35A | M800-50A | VACOFLUX 50 |
|---|---|---|---|---|
| 2 | 377.1 | 367 | 391 | 414 |
| 3 | 1000 | 973.4 | 1037 | 1098.3 |
| 4 | 1769.6 | 1722.5 | 1835.2 | 1943.6 |
| 5 | 2621.4 | 2551.6 | 2718.6 | 2879.1 |
| 6 | 3514.6 | 3421 | 3644.8 | 3860 |
| 7 | 4425.4 | 4307.6 | 4589.4 | 4860.4 |
| 8 | 5340.9 | 5198.7 | 5538.8 | 5865.9 |
| 9 | 6254.6 | 6088 | 6486.4 | 6869.4 |
| 10 | 7163.4 | 6972.7 | 7428.9 | 7867.5 |

*B.2. Lateral Vibration*

According to (5), owing to discrepancy in force values the lateral vibrations will be various.

$$[M]\begin{Bmatrix}\ddot{x}\\\ddot{y}\end{Bmatrix} + [K]\begin{Bmatrix}x\\y\end{Bmatrix} = \begin{Bmatrix}m\omega^2 e \cos \omega t\\m\omega^2 e \sin \omega t\end{Bmatrix} \quad (5)$$

Where, $M$ is mass matrix, $K$ is effective stiffness matrix, and $x, y$ are lateral vibrations (Displacements) along *x* and *y* axes, respectively [11]. According to rotary and cyclic motion in rotor, its angular velocity can be written as Fourier series. Then, by solving (5), *x* and *y* will be shown in Table 3. The discrepancy in employed soft magnetics gives various forces and subsequently lateral vibrations under same operating conditions. According to Table 3, the highest and lowest lateral vibrations in both topologies belongs to M235-35A and VACOFLUX 50, respectively. Also, the lateral vibrations of the Delta-type are higher than the V-type rotor configuration.

*B.3. Natural Frequency*

Additionally, understanding the natural frequency values and the natural modes gives the vision of physical and mechanical constraints in the design of the rotating machine system and then avoids resonance of the excitation frequency resulting from the rotation of the rotor and drive with the natural modes of the system [12], [13]. As shown in (6), the natural frequency values have a dependency on stiffness and mass.

$$f_n \propto \sqrt{K/m} \quad (6)$$

Where, $f_n$ is natural frequency. In the analysis of lateral vibrations, (6) is rewritten as (7)-(9). consequently, in a system with the same physical configuration, natural frequency values directly rely on Young's modulus and inversely on mass density [11].

$$K = EA/L \quad (7)$$
$$m = \rho V \quad (8)$$
$$f_n \propto \sqrt{E/\rho} \quad (9)$$

According to that, the physical configuration not to be effective, and only the effect of the used soft magnetics is regarded; the natural frequency values of the stator are shown in Table 5. Consequently, the higher value of the natural frequency in each mode, the higher frequency of excitation, including rotation and switching, can be chosen in the same operating conditions with more reliability. Therefore, VACOFLUX 50 is more satisfactory, and M235-35A is less attractive.

IV. CONCLUSION

In the above study, two topologies of IPMSM to employ in the AC dynamometer were designed optimally, which gives advantages, such as low $T_{ripple}$, high efficiency, high $T_{max}$, and high $T_{avg}$. By changing the soft magnetics in the rotor and stator core, the behavior of the AC dynamometer under the same operating conditions was investigated electrically and mechanically through multiphysics analysis. The investigations showed that the Hiperco 50 and VACOFLUX 50 have the highest saturation flux density, and in the case of V-type topology, Hiperco 50 has the highest $T_{max}$, $T_{avg}$, and the highest efficiency at 3000 *rpm*. Correspondingly, in the V-type topology, the lowest $T_{ripple}$ belongs to the VACOFLUX 50. In Delta-type topology, the highest $T_{max}$ and $T_{avg}$ belong to Hiperco 50 and VACOFLUX 50, the lowest $T_{ripple}$ belongs to M350-50A, and the highest efficiency belongs to M235-35A and Hiperco 50. Noteworthy The negative effect of cobalt-containing soft magnetics in Delta-type topology is $T_{ripple}$ percentage. The lowest $T_{cogging}$ in both topologies belongs to the Hiperco 50 and M800-50A. In the mechanical analysis, investigations were on average stress, lateral vibrations, and natural frequency values, and VACOFLUX 50 gives better lateral vibrations and natural frequency behavior. On the other hand, M235-35A has less stress and lower mass. A summary of the multiphysics analysis of the employed soft magnetics in the designed IPMSM is displayed in Figure 5. Among the soft magnetics category, Hiperco 50 and VACOFLUX 50 have the most cost. M235-35A and M350-50A are on the next stairs. Accordingly, electric motor designers should take a tradeoff between cost and materials behaviors, electrically and mechanically. This study shows comprehensively a soft magnetic with satisfactory behavior, electrically and mechanically, won't fit in all designs, and the status of the objectives should be investigated based on usage.

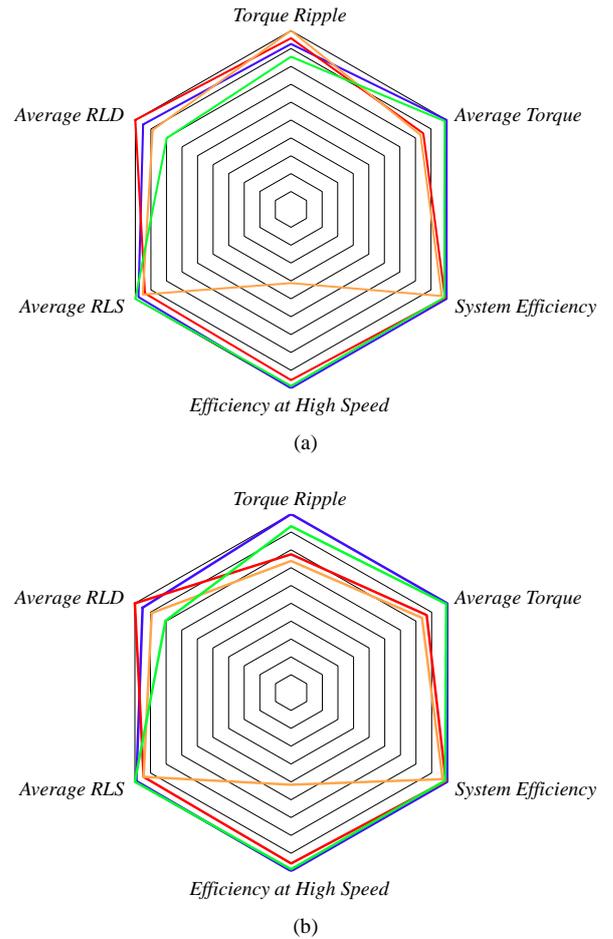

Figure 6. A summary of the multiphysics analysis of the employed soft magnetics, including Hiperco 50 (Blue), VACOFLUX 50 (Green), M235-35A (Red), and M350-50A (Orange) in the designed IPMSMs: (a) V-type, (b) Delta-type.